\documentclass[aps,twocolumn,prl,superscriptaddress,showpacs,tightenlines]{revtex4-1}
\usepackage{amsmath}
\usepackage{graphicx}
\usepackage{epsfig}
\usepackage{amsfonts}

\begin{document}

\newcommand{\nn}{\nonumber}
\newcommand{\ms}[1]{\mbox{\scriptsize #1}}
\newcommand{\dg}{^\dagger}
\newcommand{\smallfrac}[2]{\mbox{$\frac{#1}{#2}$}}
\newcommand{\la}{\langle}
\newcommand{\ra}{\rangle}
\newcommand{\ket}[1]{| {#1} \ra}
\newcommand{\bra}[1]{\la {#1} |}
\newcommand{\pfpx}[2]{\frac{\partial #1}{\partial #2}}
\newcommand{\dfdx}[2]{\frac{d #1}{d #2}}
\newcommand{\ioh}{-\frac{i}{\hbar}}
\newcommand{\ohh}{-\frac{1}{\hbar^2}}
\newcommand{\half}{\smallfrac{1}{2}}

\title{Modulated Electromechanics: Large Enhancements of Nonlinearities}
\author{Jie-Qiao Liao}
\affiliation{CEMS, RIKEN, Saitama 351-0198, Japan}
\author{Kurt Jacobs}
\affiliation{Department of Physics, University of Massachusetts at Boston, Boston, MA 02125, USA}
\affiliation{CEMS, RIKEN, Saitama 351-0198, Japan}
\author{Franco Nori}
\affiliation{CEMS, RIKEN, Saitama 351-0198, Japan}
\affiliation{Physics Department, The University of Michigan, Ann Arbor, Michigan 48109-1040, USA}
\author{Raymond W. Simmonds}
\affiliation{National Institute of Standards and Technology, 325 Broadway, Boulder, Colorado 80305, USA}

\begin{abstract}
It is well-known that the nonlinear coupling between a mechanical oscillator and a superconducting oscillator or optical cavity can be used to generate a Kerr-nonlinearity for the cavity mode. We show that the strength of this Kerr-nonlinearity, as well as the effect of the photon-pressure force can be enormously increased by modulating the strength of the nonlinear coupling. We present an electromechanical circuit in which this enhancement can be readily realized.
\end{abstract} 

\maketitle

Nano-electro-mechanical systems are superconducting circuits that are coupled to tiny, high-frequency mechanical oscillators, and can now be realized in the laboratory~\cite{Blencowe04, Poot12, Xiang13, Rocheleau10, Massel11, Teufel11a, Teufel11, Verhagen12}. The coupling between a single LC oscillator and a mechanical oscillator is nonlinear because the frequency of the LC oscillator depends on the position of the latter. This nonlinear coupling is the same as that in opto-mechanical systems, in which one of the mirrors of an optical cavity  is a mechanical oscillator~\cite{Aspelmeyer12}. Because of the nonlinear interaction nano-electro-mechanical systems posses a potentially rich dynamics and provide a fertile field for the realization of quantum control~\cite{Mancini97, Bose97, Rabl11, Nunnenkamp11}. The main obstacle to realizing the quantum effects of the nonlinear interaction is that these effects are much weaker (slower) than the strength (or rate) of the nonlinear interaction itself. If we denote the latter by $g$ (defined precisely below), then the rate of the induced Kerr nonlinearity is ($g^2/\Omega$), where $\Omega$ is the mechanical frequency, and the displacement of the mechanical resonator induced by the photons is proportional to $g/\Omega$~\cite{Mancini97, Bose97}. Since in present systems $g \ll \Omega$, the nonlinear effects are negligible. 

Here we show that the Kerr nonlinearity and the displacement induced by the photon-pressure force can be enhanced by a factor of $\Omega/g \sim 10^2 \mbox{-}10^4$ by modulating $g$ at a frequency close to the mechanical frequency. This enhancement can be understood by viewing the nonlinear interaction as off-resonant, and the modulation as bringing it onto resonance. We also present evidence from numerical optimization that the rate $g$ is the maximum rate that can be achieved for the effective Kerr nonlinearity. This nonlinearity can be used to prepare nonclassical states of the cavity mode as well as the mechanics, and the optical force can be used to realize opto-mechanical entanglement at the single-photon level, as a transducer for measuring displacement or photon-number, and to probe foundational questions in quantum theory~\cite{Bose99, Marshall03}. We note that a number of works have recently shown that the optomechanical interaction can induce various nonlinear effects at the rate $g$ if two optical modes are coupled together, one of which interacts with the mechanics~\cite{Stannigel12, Ludwig12, Lu13}. We show that the enhancement induced by these two-mode techniques can also be viewed as a result of bringing the optomechanical interaction into resonance.  

We begin by showing how the modulation of $g$ enhances the Kerr nonlinearity and the effect of the photon-pressure force, and discuss how the enhancement of nonlinearities in two-mode systems is explained by the same mechanism. We then consider two ways in which the modulation of $g$ could be realized in electro-mechanical systems. Using the second of these methods we present a readily realizable circuit in which the effect of the photon-pressure force is enhanced by more than three orders of magnitude. 

The electro-mechanical coupling is given by the Hamiltonian~\cite{Law95}
\begin{equation}
  H = \hbar \omega a^\dagger a + \hbar \Omega b^\dagger b + \hbar g a^\dagger a (b + b^\dagger) , \label{eq1}
\end{equation} 
where $a$ and $b$ are the annihilation operators for the superconducting and mechanical modes, respectively, $\omega$ and $\Omega$ are their respective frequencies, and $g$ is the elecro-mechanical coupling rate. The unitary evolution operator generated by this Hamiltonian, $U(t) = e^{-i H t/\hbar}$ can be written in the following simple form~\cite{Mancini97, Bose97}: 
\begin{eqnarray}
U(t) &=&\exp(-i\omega  a^\dagger a t) \exp \left[ i \mu (a^\dagger a)^2 \right] \nonumber\\
  & &  \times  \exp[- i a^{\dagger}a \sqrt{2} ( \lambda_x x -\lambda_p p)]  \exp(-i\Omega b^{\dagger}b t) , \;\;
\end{eqnarray}
with $\lambda_x = (g/\Omega)\sin(\Omega t)$, $\lambda_p = (g/\Omega)[1-\cos(\Omega t)]$, and 
\begin{equation}
   \mu =  (g^2/\Omega) \left[ t  - \sin(\Omega t)/\Omega  \right] , 
\end{equation} 
and we have defined the dimensionless mechanical position and momentum operators by $x \equiv (b + b^\dagger)/\sqrt{2}$ and $p \equiv -i(b - b^\dagger)/\sqrt{2}$. Two key effects can be read off from $U$. The first is that the electrical mode displaces the mechanical mode by the (dimensionless) phase-space distance $\Delta s \equiv \sqrt{\Delta x^2 + \Delta p^2} = \sqrt{8} (g/\Omega) n$, at time $t = \pi/\Omega$, where $n$ is the number of photons in the electrical mode. The second effect is that at times  $\tau = 2\pi m/\Omega$, for integer $m$, the electrical mode undergoes the evolution  
\begin{equation}
   U(\tau) = \exp\left[ -i \left( \omega  a^\dagger a + \Omega b^{\dagger}b - (g^2/\Omega) (a^\dagger a)^2  \right) \tau \right] ,  
\end{equation} 
where the size of the effective Kerr-nonlinearity is $\chi = g^2/\Omega$. Both the displacement of the mechanical mode, and the Kerr-evolution contain the small factor $g/\Omega$. If we examine the optomechanical $H$ above, and move into the interaction picture, then the interaction is $H_{\ms{I}} = \hbar g a^\dagger a (b e^{-i\Omega t} + b e^{i\Omega t})$. Both terms oscillate at the mechanical frequency $\Omega$, and are therefore off-resonant for $g \ll \Omega$. As a result the rotating-wave approximation for $g \ll \Omega$ eliminates the interaction. 

If we modulate the interaction rate $g$ so that $g \rightarrow \tilde{g}(t) = g \cos(\nu t)$, with $\nu = \Omega - \delta$, then we can bring the interaction near to resonance, with the remaining detuning equal to $\delta$. If we then move into the interaction picture with respect to the Hamiltonian $H(\nu) = \hbar \nu b^\dagger b$, and make the rotating-wave approximation ($g \ll \nu$), the effective Hamiltonian for the joint system is that given by $H$ but with $\Omega$ replaced by $\delta$ and $g$ replaced by $g/2$. By choosing $\delta = g/2$, the rate of the Kerr nonlinearity becomes $\chi = g^2/(4\delta) = g/2 \gg g^2/\Omega$, and the mechanical displacement is similarly $\Delta s = \sqrt{8}n \gg \sqrt{8} (g/\Omega) n$. 

The above magnification of the nonlinear effects is potentially very large, but given the dependence of these rates on $\delta$ we might wonder whether they could be made arbitrarily large by reducing $\delta$ further. This is not true for the Kerr term, as we now show. We can certainly set $\delta = g/(2r)$ so that the Kerr term in the evolution operator $U$ has $\chi = r g/2$. The catch is that we must wait for the two oscillators to decouple, and this takes the minimum time $\tau = r 4\pi /g$, which increases with $r$. As an example, if we want to use the Kerr term to prepare the superposition $|\mbox{cat}(\alpha)\rangle \equiv (|\alpha\rangle - i |-\alpha\rangle )/\sqrt{2}$, where $|\pm\alpha\rangle$ denotes a coherent state with complex amplitude $\pm\alpha$, then we need $\chi \tau = \pi/2$. If we choose $r$ to minimize the time taken, then the minimum is at $r = 1/2$, the effective value of $\chi$ is $g/4$, and the time taken to prepare the cat state is $\tau = 2\pi/g$. 

The above analysis does not exclude the possibility that a shorter time might be obtained by allowing $g$ to be an arbitrary function of time. To answer this question we perform a numerical search for time-dependent control strategies (that is, ways to change $g$ and $\Omega$ with time) to generate the unitary $V = \exp[i \pi (a^\dagger a)^2/2]$ in the minimum time. If we can prepare $|\mbox{cat}(\alpha)\rangle $ in time $\tau$, then the realizable Kerr rate is $\chi = \pi/(2\tau)$. To do this we divide the time interval $[0,\tau]$ up into $N$ segments, and allow $g$ and $\Omega$ to take a different value on each segment. We then perform a gradient search to find an optimal set of values for $g$ and $\Omega$, given a maximum value for $g$. Since the system consists of two oscillators the state space is potentially large. Fortunately, the problem allows a simplification: the Hamiltonian commutes with $a^\dagger a$, and so preserves the populations of the number states. This means that we loose no accuracy in truncating the LC oscillator in the number basis. The mechanical oscillator on the other hand requires a much larger state-space, because the evolution generates coherent states for this oscillator. To perform the numerical optimization we use just three number states for the superconducting resonator, thirty for the mechanics, and set $g = \pi$. We choose an arbitrary initial state for the LC-oscillator, $|\psi (0)\rangle$, and use the gradient search to maximize the fidelity between $|\psi(\tau)\rangle$ and the desired final final state $V |\psi(0)\rangle$~\footnote{The fidelity between two density matrices $\rho$ and $\sigma$ is $F = \mbox{Tr}[\sqrt{\sigma^{1/2}\rho\sigma^{1/2}}]$. When the two states are pure this reduces to the absolute value of their inner product~\cite{Jozsa95}}, for a range of values of $\tau$. We also run the optimization with two values of $N$ ($N=10$ and $N=15$) to ensure that $N$ does not limit the fidelity. We find that for $\tau \geq 1$ we can always obtain a fidelity equal to unity, with essentially arbitrary accuracy. As soon as we set $\tau < 1$ this is no longer possible. If we define our figure of merit as $\varepsilon = 1 - F$, where $F$ is the fidelity, then for $\tau = 0.99, 0.95, 0.9, 0.8$ we obtain $\varepsilon = 1.1\times 10^{-4}, 2.7\times 10^{-3}, 1\times 10^{-2}, 3.6\times 10^{-2}$. This clear change in behavior around $\tau = 1$ gives us considerable confidence that $\chi = g/2$ is the maximum Kerr rate that can be generated with this system. 

The situation regarding the photon-pressure force is a little different. In this case, the interaction is linear driving from the point of view of the mechanics, but by a force proportional to the photon number $n$. To make the most of this force we should drive the mechanical oscillator at its resonance, and we do this by choosing $\delta = 0$. The resulting evolution of the (dimensionless) mechanical momentum operator, in the interaction picture, is   
\begin{equation}
  p(t) = p(0) + \sqrt{2} (g/2) a^\dagger a t . 
\end{equation}
The upper limit to the phase-space displacement is now only that imposed by the damping rate of the mechanics, $\gamma$, and this is 
\begin{equation} 
  \Delta s_{\ms{max}} = \max [p(t) - p(0)] = \frac{gn}{\sqrt{2}\gamma}  .   
  \label{eq6}
\end{equation} 
where $n$ is the number of photons in the resonator~\footnote{Equation (6) assumes the resonator has reached its steady-state for a given photon number. We note that since the damping of the LC-oscillator is faster than the mechanics, obtaining the steady-state for a fixed  $n$ would require repeated re-initialization of the LC oscillator in the number state $n$.}.  Note that the photon force generates a coherent state of the resonator. If the oscillator starts in the vacuum state, then this coherent state is $|\beta(t)\rangle$, with $\beta (t) = -i(g/2) n t$. The average number of phonons in the coherent state is then $|\beta|^2 = (gnt/2)^2$, and the steady-state value is $|\beta|_{\ms{ss}}^2 = (gn/\gamma)^2$. 

Very recently it has been shown that if a mechanical resonator is coupled to two optical (or superconducting) modes, effective nonlinearities can be realized that are much stronger than with a single mode~\cite{Stannigel12, Ludwig12, Lu13}. In Refs.~\cite{Stannigel12, Ludwig12} the two optical modes are coupled together via a linear interaction, and their coupling frequency is chosen to be half the mechanical frequency. One can then determine the ``normal'' modes generated by the linear interaction, being the linear combinations of the optical modes that are uncoupled. The frequency difference between these modes is twice the coupling rate, and thus equal to the mechanical frequency. The result is that there is now a resonant (three-way) interaction between the normal modes and the mechanics at rate $g$, because the frequency of one normal mode is equal to the sum of the mechanical frequency and that of the other normal mode. Another way to think of this is that the first normal mode and the mechanics are not resonant, but the addition of the second normal mode makes up the frequency (energy) difference and allows them to interact resonantly. In our modulation technique, the second normal mode is effectively replaced by a classical drive.  

In the second two-mode scheme, presented in~\cite{Lu13}, the mechanics is coupled to one optical cavity mode and one superconducting mode. This time it is the mechanical and superconducting modes that are coupled, and the resulting normal modes determined. Both of these normal modes now interact with the optical mode via the nonlinear optomechanical coupling. The mechanical and superconducting frequencies are chosen so that one of the normal modes has a much lower frequency, equal to the optomechanical coupling rate $g$, and as a result the nonlinear interaction is close to resonance. While the previous two-mode scheme essentially upconverts the optomechanical coupling to match the mechanical frequency, the second effectively does the opposite, reducing the mechanical frequency to match the rate of the former. 

We now turn to the question of how $g$ might be modulated in real electromechanical circuits. Note that to obtain the largest nonlinear enhancement, for a given value of the interaction rate $g$, the amplitude of the modulation would need to be equal to $g \equiv g_{\ms{max}}$. If this amplitude is instead some fraction $\eta$ of $g_{\ms{max}}$, so that the modulated interaction rate is $g(t) = g_{\ms{max}} [(1- \eta) + \eta \cos(\nu t)]$, then the resulting Kerr rate and maximum displacement are simply reduced by the same faction: $\chi = \eta g_{\ms{max}} /4$, $\Delta s_{\ms{max}} = \sqrt{2} \eta (g_{\ms{max}}/\gamma) n$. 

\begin{figure} 
\leavevmode\includegraphics[width=0.9\hsize]{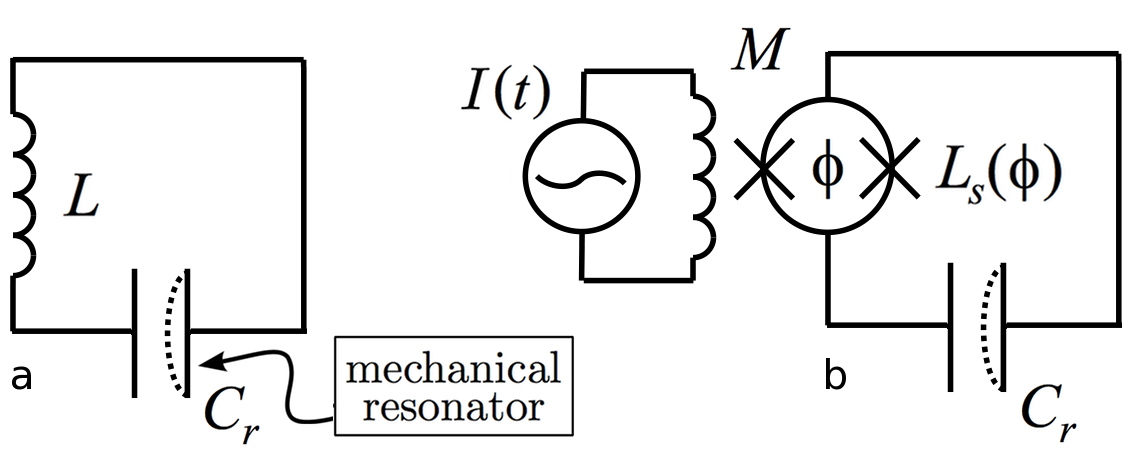}
\caption{(a) A simple LC-circuit capacitively coupled to a mechanical resonator. The resonator forms one of the plates of the capacitor. (b) An electromechanical circuit in which the nonlinear coupling rate between the superconducting (LC) oscillator and the mechanical oscillator can be modulated. The circuit loop on the right is effectively an LC oscillator in which the inductance is determined by $\phi$, the flux applied between two Josephson junctions. This flux is created by the circuit loop on the left. The mechanical resonator forms one of the plates of the capacitor in the usual capacitative coupling configuration.} 
\label{fig1} 
\end{figure}

Consider a superconducting LC oscillator capacitively coupled to a mechanical resonator, the circuit for which is shown in Fig.~\ref{fig1}. The mechanical oscillator forms one plate of the capacitor, and thus changes the capacitance as it moves. The Hamiltonian for the circuit is given by Eq.(\ref{eq1}), where the frequency of the LC oscillator is $\omega = 1/\sqrt{LC_r}$ with $L$ the inductance and $C_r$ the capacitance, and the nonlinear coupling rate is $g = \omega/(2d)\sqrt{\hbar/(2 m \Omega)}$, with $d$ the distance between the capacitor plates, and $m$ the mass of the mechanical oscillator. It is therefore possible to modulate $g$ by modulating $\omega$ or modulating $d$. In fact, modulating $d$ also modulates $\omega$ because it modulates the capacitance. So long as the frequency of the modulation is small compared to the frequency, $\omega$, of the LC oscillator, the populations of the energy eigenstates of this oscillator (the photons) will adiabatically follow the change in $\omega$. It is because of this adiabatic following that the only change to the Hamiltonian induced by the modulation is the replacement of $\omega$ and $g$ by their time-dependent values.    

One way to modulate the electrical frequency is to add another capacitor in parallel to $C_r$, and modulate the distance between its plates. In fact, this method has very recently been suggested as a way to modulate a capacitance to provide a linear coupling between the motion of a trapped ion and an LC-oscillator~\cite{Kielpinski12}. There the authors suggest the use of a bulk-acoustic wave generator to modulate the second capacitance. Using a 1 GHz LC-oscillator, they obtain a modulation amplitude corresponding to $\eta \approx 0.2$. If we have a 10 MHz mechanical resonator with electro-mechanical coupling rate $g = 2\pi \times100$ Hz~\cite{Palomaki13}, and modulate near 10 MHz, we would achieve a Kerr nonlinearity with rate $\chi = \eta g/4 = 10\pi~\mbox{s}^{-1}$. This is an increase in the Kerr rate by a factor of $500$. Of course, we can create strong Kerr nonlinearities for superconducting oscillators using Josephson-junctions~\cite{Kirchmair13}, but these nonlinear oscillators saturate at a few hundred photons. If electro-mechanical coupling rates continue to increase, this technique could be used to produce mesoscopic superpositions in linear LC-resonators, with much larger photon numbers. 

While the use of an acoustic wave generator to modulate the capacitance may well be possible, such a system has not yet be realized experimentally. We now consider a circuit in which we can modulate $g$ by using circuit elements that have already been employed and well-tested in electro-mechanical circuits. Instead of modulating the capacitance, we replace the inductor with two Josephson-junctions (JJs) which allows us to modulate the inductance. In this case the purpose is not to generate a Kerr nonlinearity, since the JJs do this themselves. Rather we wish to maximize the amount by which the photons can affect the  mechanical motion. 

The circuit we propose is shown in Fig.~\ref{fig1}(b). The circuit loop on the right can be thought of as a standard LC oscillator in which the inductor has been replaced by a pair of Josephson junctions in parallel. We chose the parameters of the JJs so that they are only weakly nonlinear, and thus effectively provide a linear inductance. The effective inductance, $L_{\mbox{\scriptsize s}}$, produced by the pair of JJs depends on the external flux, $\phi$, applied between them~\cite{Zakka11}, and the sole purpose of the loop on the left is to apply this flux. The resulting inductance is $L_{\mbox{\scriptsize s}}(\phi) = L_{\mbox{\scriptsize J}}/\cos(\pi\phi) = \phi_0/[4\pi I_0 \cos(\pi\phi)]$, where $\phi_0 = h/(2e)$ is the flux quantum, and $I_0$ is the critical current of each of the junctions. The frequency of the LC oscillator is $\omega = 1/\sqrt{L_{\mbox{\scriptsize s}} C_{\mbox{\scriptsize r}}} = \omega_{\mbox{\scriptsize max}} \sqrt{\cos(\pi\phi)}$, with $\omega_{\mbox{\scriptsize max}} = \sqrt{4\pi I_0/ \phi_0 C}$. 

The coupling between the LC oscillator and the mechanical oscillator stems from the fact that the capacitance is inversely proportional to the distance between its plates. If $x$ is the position of the mechanical oscillator, $C$ is the value of $C_{\mbox{\scriptsize r}}$ when $x=0$, and the amplitude of the oscillation is small compared to the distance between the plates, then $C_{\mbox{\scriptsize r}} \approx C (1 - x/d)$. Including this in the expression for the frequency of the LC oscillator we obtain $\omega = \omega_{\mbox{\scriptsize max}} \sqrt{\cos(\pi\phi)} [ 1 + x/(2d) ]$. 
The full Hamiltonian for the two oscillators is given by substituting this expression for $\omega$ into the Hamiltonian for the non-interacting oscillators, $H_0 = \hbar \omega a^\dagger a + \hbar \Omega b^\dagger b$. The resulting Hamiltonian is that given in Eq.(\ref{eq1}) with 
\begin{equation}
    \omega = \omega_{\mbox{\scriptsize max}} \sqrt{\cos(\pi\phi)} , \;\;\;  g =  g_{\mbox{\scriptsize max}} \sqrt{\cos(\pi\phi)} ,  
\end{equation} 
where $g_{\mbox{\scriptsize max}} = \omega_{\mbox{\scriptsize max}} x_{\mbox{\scriptsize zp}} /(2d)$, $x_{\mbox{\scriptsize zp}} = \sqrt{\hbar/(2 m \Omega)}$ is the ``zero-point motion'' of the mechanics, and we have used the fact that $x = x_{\mbox{\scriptsize zp}}(b + b^\dagger)$.  If we vary $\phi$ with time, and ensure that the rate of change of $\phi$ is small compared to $\omega$, then the adiabatic approximation preserves the state of the system with respect to the eigenvectors of the changing mode operators. The result is that the mode operator $a$ is preserved, and it is merely $\omega$ and $g$ that change with time. By varying $\phi$ we can choose $g$ to be any function of time, within the constraint $0 < g < g_{\mbox{\scriptsize max}}$. 

To obtain the modulation $\tilde{g} = g \cos(\nu t)$ we can choose $\sqrt{\cos(\pi\phi)} = [1 + \cos(\nu t)]/2$, and this gives $\tilde{g} = (g_{\mbox{\scriptsize max}}/2)[1 + \cos(\nu t)]$, and thus $\eta = 1/2$. The rate at which the optical force increases the momentum of the mirror is then $g_{\mbox{\scriptsize max}} n /\sqrt{8}$, with $n$ the number of photons. So at what rate could the momentum be changed with present technology? The largest value of $g$ that has been achieved to-date is $g_{\mbox{\scriptsize max}} = 2\pi \times 230~\mbox{Hz} = 1445~\mbox{s}^{-1}$~\cite{Teufel11}. With the same technique we estimate that increasing $g$ by a factor of $4$ is feasible, giving $g = 5780~\mbox{s}^{-1}$. Using a single qubit to load photons into the LC resonator, the preparation of a number-state with $n = 10$ or even $n = 50$ is entirely feasible~\cite{Hofheinz09}. With $n=10$ and the above value for $g$, the time taken to displace the mechanical oscillator by an average of 25 phonons is $t = 1/g \sim 180 \mu\mbox{s}$. The average steady-state displacement, for 10 MHz mechanics with a quality factor of $10^5$ is $\sim 90$ phonons. 

To summarize, we have shown that modulating the electromechanical coupling rate increases the Kerr nonlinearity and the effect of the photon-pressure force by orders of magnitude, and we have shown how this modulation can be realized. Potential uses include the generation of high-amplitude nonclassical states, observing the mechanical effects of quantum states of light, and realizing non-demolition measurements of photons. It is interesting to note that a similar enhancement could be achieved in opto-mechanical systems if a feasible method were found to  perform the modulation. 

\emph{Acknowledgements} JQL is supported by a Foreign Postdoctoral Fellowship (No.\ P12503) of the Japan Society for the Promotion of Science (JSPS). KJ is partially supported by the NSF projects PHY-1005571 and PHY-1212413, and the ARO MURI grant W911NF-11-1-0268. FN is partially supported by the ARO, JSPS-RFBR contract No. 12-02-92100, Grant-in-Aid for Scientific Research (S), MEXT Kakenhi on Quantum Cybernetics, and the JSPS via its FIRST program. 

\vspace{-3mm}


%

\end{document}